# Background radioactivity of construction materials, raw substance and ready-made CaMoO$_4$ crystals

O.A. Busanov[1], R.A. Etezov[2], Yu.M. Gavriljuk[2], A.M. Gezhaev[2], V.V. Kazalov[2,a], V.N. Kornoukhov[1,3], V.V. Kuzminov[2], P.S. Moseev[1], S.I. Panasenko[2,4], S.S. Ratkevich[2,4], S.P. Yakimenko[2]

[1]OJSC "Fomos Materials", Moscow, Russia
[2]Institute for Nuclear Research of RAS, Moscow, Russia
[3]Institute for Theoretical and Experimental Physics, Moscow, Russia
[4]Karazin Kharkiv National University, Kharkiv, Ukraine

**Abstract.** The results of measurements of natural radioactive isotopes content in different source materials of natural and enriched composition used for CaMoO$_4$ scintillation crystal growing are presented. The crystals are to be used in the experiment to search for neutrinoless double beta decay of $^{100}$Mo.

## 1 Introduction

International collaboration AMoRE (Advanced Mo based Rare process Experiment) is planning to use isotopically enriched $^{40}$Ca$^{100}$MoO$_4$ monocrystals as cryogenic scintillation detector for studying the process of neutrinoless double beta decay of the isotope $^{100}$Mo (Q$_{\beta\beta}$=3034 keV) [1]. Experimental detection of this process can be the proof of the Majorana nature of neutrinos and will assess the value of the effective mass of the particle. A simultaneous registration of phonon and scintillation signals will be used to suppress internal background [2].

The main parameter which limits the sensitivity of the experiment is the content of radioactive isotopes in the crystals material. The contents should be monitored in all components included in the charge from which the crystals were grown, in the charge itself and in the ready-made crystals. Such control allows to select the most pure raw materials and to make a technological chain of manufacture of crystals providing the highest radioactive purity of the product. As part of this work, a number of measurements of radioactive purity of the raw materials for the manufacture of a charge, the charges and ready-made crystals were carried out with the low-background spectrometer with ultrapure HPGe-detectors located in underground laboratories at the Baksan Neutrino Observatory of INR RAS [3].

## 2 NIKA facility description

The Low-background Chamber (NIKA) is placed at a depth of 660 m of water equivalent (m w.e.), where a cosmic ray background is reduced by ~2000 times. Its walls are covered by low radiation concrete with the thickness of 50 cm on a dunite base. Dunite is an ultrabasic rock with the low content of natural radioactive isotopes. An inner volume of the workroom (4×4×3 m$^3$) is separated from the ceiling, floor and concrete walls by the 50 cm dunite layer. Such construction enabled to lower a gamma-background by 200 times comparably to the empty cavity. The low-background shield is placed in the workroom. It consists of 8 cm borated polyethylene + 23 cm Pb + 12 cm Cu. There is a rectangular cavity with the dimensions of 30×30×30 cm$^3$ in the center of Cu protection layer. The heads of three ultra low-background semiconductor germanium detectors are placed in this cavity. Detectors placement geometry in working cavity and their numeration are shown in Fig.1. Mass of each detector is about 1 kg as it is shown in Table 1. Protective cases of the detectors are made from a high purity electrolytic copper. The detectors are cooled through the cold fingers took out from the shield into the Dewar vessel. A vapor of liquid nitrogen outgoing from the Dewar is used for the cavity blow. The detectors "1,2,4" in Table 1 are made of the high-purity germanium, enriched with $^{76}$Ge isotope to 87%. Detector "3" is made of the high-purity natural germanium. The detector #3 is currently used for a radioactive purity examination of material samples in the NIKA-spectrometer. The de-tector #4 is used in the SNEG-spectrometer described below.

[a] Corresponding author: vvk1982@mail.ru

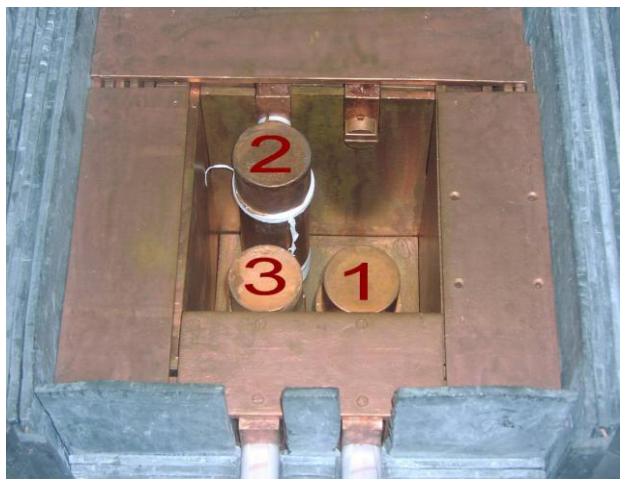

**Figure 1.** Ultra low background germanium γ-spectrometer.

**Table 1.** Parameters of the semiconductor detectors.

| The detector's number → <br> Parameters ↓ | #1 | #2 | #3 | #4 |
|---|---|---|---|---|
| Material | Ge-76 | Ge-76 | Ge-nat | Ge-76 |
| Total mass, g | 1006 | 896 | 1056 | 968 |
| Effective mass, g | 630 | 680 | 980 | 642 |

## 2.1 The recording equipment

Pulses from the detector #3 feed an amplification unit which consists of charge sensitive preamplifier and shaping amplifier with 4 μs integration and differentiation shaping time. Output shaped pulses are recorded with two-channel digital integration and differentiation shaping time. Output shaped pulses are recorded with two-channel digital oscilloscope board La-n20-12PCI of Rudnev-Shilyaev firm, built-in personal computer. The pulses are digitized with a frequency of 6.25 MHz and are written to the hard disk for later analysis. After accumulating the required amount of data, a collection is completed and the data are processed on a separate PC. The total pulse shape is analyzed to construct an amplitude spectrum of the integrated pulse areas. The pulses which have non ionization nature such as surface micro-discharges and breakdowns are excluded because their shapes have specific features.

## 3 SNEG facility description

This low-background facility is based on a low-background HPGe-detector #4. General view of an equipment of the SNEG-spectrometer is shown in Fig 2.

This spectrometer was created in a new Deep Underground Low-Background laboratory at a depth of 4900 m w.e. (DULB-4900) [4].

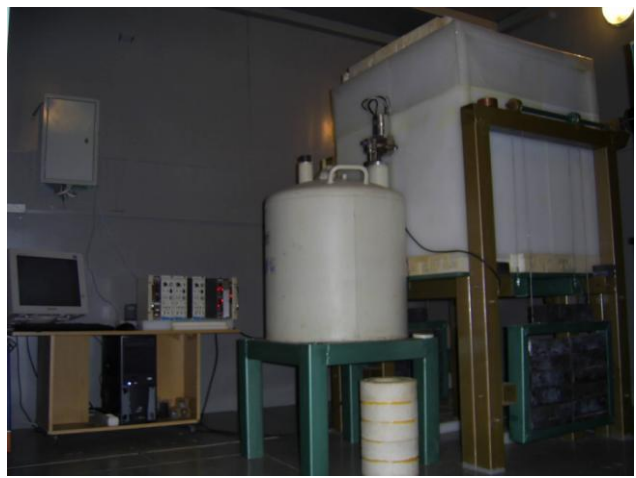

**Figure 2.** General view of the SNEG-spectrometer.

In the center of the photo one can see a stand with the 80 l Dewar vessel used to cool off the vertical part of the cold finger of the HPGe detector #4. The head of the detector is inside the shield which consists of 5-8 cm polyethylene + 1mm Cd + 18 cm Pb + 15 cm Cu. A cavity for placement of the samples is placed around the head end. The geometry of the cavity can be changed by filling a part of the volume by the high-purity copper cylinder. The shield of the detector head is visible to the right of Fig. 2.

Samples are loaded into the working volume through the bottom entrance. A part of the bottom shield is mounted on the lifted platform. To install a sample, the platform is lowered to the floor. A sample is placed in a special recess of the movable upper copper block of the shield. The platform is raised. In this case, shield is closed and the sample is below the head of the detector. A residual cavity volume in the shield is purged by vapor of the liquid nitrogen exiting the Dewar to remove the radioactive $^{222}$Rn from the working region. Vapor feeds by means of a copper pipe laying in the upper channel parallel to the cold finger and goes freely through the slots in the bottom of the shield. The upper part of the copper layer of the shield is sealed with a plastic film and a sealant.

Electronic equipment and a processing procedure are similar to that used in the NIKA-spectrometer.

## 4 Results of measurements

Table 2 provides a complete list of the examined samples, their chemical and technical characteristics as well as the results of the determination of the specific activity of radioactive isotopes. The samples in the first column are placed in a sequence of a creation. The numbers in square brackets in column for $^{208}$Tl specified equilibrium activity of $^{232}$Th. An indicated time of measurements (T) corresponds to the accumulation of useful information at a constant counting rate achieved after blowing of the working volume from the $^{222}$Rn and its daughter isotopes and decay of non-equilibrium short lived radioactivity in packages of samples.

**Table 2.** Activity of radioactive isotopes in the samples [Bq/kg] (limits are at 95% C.L.)

| Sample, material | Isotopes | | | |
|---|---|---|---|---|
| | $^{40}$K | $^{228}$Ac=($^{232}$Th) | $^{208}$Tl [($^{232}$Th)] | $^{214}$Bi=($^{238}$U) |
| | Activity of radioactive isotopes | | | |
| 1. Charge A1 ~CaMoO$_4$ Mass=414 g, T=115 h | (2.7±0.5)·10$^{-1}$ | (5.3±1.9)·10$^{-2}$ | (1.3±0.3)·10$^{-2}$ [(3.6±0.8)·10$^{-2}$] | 3.07±0.05 |
| 2. Charge A2 ~CaMoO$_4$ Mass=420 g, T=209 h | (4.7±0.5)·10$^{-1}$ | (2.6±1.4)·10$^{-2}$ | (1.9±0.3)·10$^{-2}$ [(5.3±0.8)·10$^{-2}$] | 3.51±0.04 |
| 3. Charge B ~CaMoO$_4$ Mass=523 g, T=143 h | (3.6±0.4)·10$^{-2}$ | ≤7.1·10$^{-3}$ | (2.8±1.5)·10$^{-3}$ [(7.8±4.2)·10$^{-3}$] | (7.8±1.1)·10$^{-2}$ |
| 4. Monocrystal CaMoO$_4$ from B Mass=348.5 g, T=498 h | (2.0±0.8)·10$^{-2}$ | ≤5.6·10$^{-3}$ | ≤1.4·10$^{-3}$ [≤3.9·10$^{-3}$] | ≤2.2·10$^{-3}$ |
| 5. Calcium formate Ca(HCOO)$_2$ Mass=400 g, T=333 h | ≤7.0·10$^{-3}$ | ≤3.0·10$^{-3}$ | ≤8.9·10$^{-4}$ [≤2.5·10$^{-3}$] | ≤1.7·10$^{-3}$ |
| 6. $^{100}$MoO$_3$ Mass=223 g, T=840 h | (5.3±0,8)·10$^{-2}$ | ≤3.8·10$^{-3}$ | ≤1.0·10$^{-3}$ [≤2.8·10$^{-3}$] | ≤2.3·10$^{-3}$ |
| 7. $^{40}$CaCo$_3$ Mass=60.25 g, T=557 h | (7.3±3.1)·10$^{-2}$ | (1.6±0.2)·10$^{-1}$ | (4.4±3.6)·10$^{-3}$ [(1.2±1.0)·10$^{-2}$] | (2.6±0.2)·10$^{-1}$ |
| 8. Monocrystal CaMoO$_4$ Mass=553.5 g, T=795 h | ≤8.2·10$^{-3}$ | ≤3.1·10$^{-3}$ | ≤6.6·10$^{-4}$ [≤1.8·10$^{-3}$] | ≤3.2·10$^{-3}$ |
| 9. $^{40}$CaCo$_3$ Mass=525.04 g, T=542 h | ≤1.24·10$^{-2}$ | (5.9±0.5)·10$^{-2}$ | (1.1±0.1)·10$^{-2}$ [(3.0±0.3)·10$^{-2}$] | (1.7±0.1)·10$^{-1}$ |
| 10. Monocrystal CaMoO$_4$ Mass=494.87 g, T=893 h | (1.2±0.6)·10$^{-2}$ | ≤3.3·10$^{-3}$ | ≤7.8·10$^{-4}$ [≤2.2·10$^{-3}$] | (1.6±0.2)·10$^{-2}$ |
| 11. $^{40}$CaCo$_3$ Mass=540 g, T=277 h | (1.4±0.5)·10$^{-2}$ | ≤3.2·10$^{-3}$ | ≤9.0·10$^{-4}$ [≤2.5·10$^{-3}$] | (5.7±0.3)·10$^{-2}$ |
| 12. ZnO Mass=1018 g, T=709 h | (3.8 ± 3.3)·10$^{-3}$ | (2.7 ± 0.8)·10$^{-3}$ | (3.2 ± 1.6)·10$^{-4}$ | (1.4 ± 0.8)·10$^{-3}$ |
| 13. Ceramic ring ZrO$_2$ Mass=720 g, T=1 h | ≤ 13 | 47 ± 4 | 15 ± 1 | 435 ± 6 |
| 14. Monocrystal CaMoO$_4$ Mass=494.87 g, T=515 h | ≤1.2· 10$^{-2}$ | ≤2.4· 10$^{-3}$ | (9.4 ± 5.4)·10$^{-4}$ | (9.5 ± 2.2)·10$^{-3}$ |
| 15. Nb$_2$O$_5$ Mass=60.78 g, T=638 h | ≤3.6·10$^{-2}$ | ≤6.8·10$^{-3}$ | ≤5.4·10$^{-3}$ | ≤5.8·10$^{-3}$ |
| 16. Monocrystal CaMoO$_4$ Mass=657.83 g, T=323 h | ≤1.4·10$^{-2}$ | (2.3±1.9)·10$^{-4}$ | (9.5±6.4)·10$^{-4}$ | (1.0 ± 0.3)· 10$^{-2}$ |
| 17. Monocrystal CaMoO$_4$ Mass=661.28 g, T=482 h | ≤1.2·10$^{-2}$ | ≤3.1·10$^{-3}$ | (1.0±0.5)·10$^{-3}$ | (1.0±0.3)·10$^{-2}$ |
| 18. Calcium formate Ca(HCOO)$_2$ (unpurified) Mass=500 g, T=543 h | ≤3.4·10$^{-2}$ | ≤9.1·10$^{-3}$ | ≤8.3·10$^{-3}$ | (5.9±3.8)·10$^{-3}$ |
| 19. Calcium formate Ca(HCOO)$_2$ (purified) Mass=503 g, T=437 h | ≤8.6·10$^{-3}$ | (1.3±1.1)·10$^{-3}$ | ≤1.3·10$^{-3}$ | (1.4±0.9)·10$^{-3}$ |
| 20. Charge CaMoO$_4$ Mass=500 g, T=380 h | ≤9.4·10$^{-3}$ | (1.9±1.3)·10$^{-3}$ | ≤1.1·10$^{-3}$ | ≤1.6·10$^{-3}$ |
| 21. Monocrystal CaMoO$_4$ cut from one side Mass=473.9 g, T=527 h | ≤1.3·10$^{-2}$ | ≤3.4·10$^{-3}$ | (5.0±4.9)·10$^{-4}$ | ≤5.3·10$^{-3}$ |

## 5 Discussion

It is seen from Table 2, that values of the $^{232}$Th activities, obtained from the activities of $^{228}$Ac and $^{208}$Tl, coincide within the bounds of the achieved statistical accuracy for the samples #1 - #6. But the $^{232}$Th activity of sample #7 obtained from the $^{228}$Ac gamma-line counting rate exceeds by ~10 times the one obtained from the $^{208}$Tl gamma-line counting rate. It means that the thorium mother isotope $^{232}$Th and daughter isotope $^{228}$Th were removed mainly from the material in a process of a preliminary cleaning. One could expect a rise of $^{208}$Tl-decay background with time due to a restoration of the equilibrium between $^{228}$Ra and $^{228}$Th.

A comparison of radioactivity levels of the charges #1 - #3 produced from the raw materials subjected to different preliminary purifications shows that a correct choice of a purification procedure allows one to reach essential decreasing of the radioisotope content.

A crystal #4 was made from a charge #3. As it is seen from Table 2, a further decreasing of the radioisotope content takes place in a process of the crystal's growth. It allows to achieve a decreasing of $^{40}$K content by ≥10 times, $^{232}$Th content by ≥2 times and $^{238}$U content by ≥30 times.

It was shown for the samples #5, #18, #19 that very high level of radioactive purity could be achieved if the calcium is purified as a calcium formate.

The sample #21 is a ready-made crystal of $CaMoO_4$. As it seen from the Table 2, this crystal has very low radioactive contaminations by $^{40}$K, $^{232}$Th and $^{238}$U which are limited by the spectrometer sensitivities at the levels of ≤1.3·10$^{-2}$ Bq/kg, ≤3.4·10$^{-3}$ Bq/kg, ≤5.3·10$^{-3}$ Bq/kg, respectively.

## 6 Conclusion

As it follows from the above consideration, a presently achieved level of industrial technology allows one to provide the experimentalists by the scintillation crystals of $CaMoO_4$ with very low radioactive isotopes contaminations.